\documentstyle[11pt]{article}
\newcommand{\be}{\begin{equation}}
\newcommand{\ee}{\end{equation}}
\newcommand{\ba}{\begin{eqnarray}}
\newcommand{\ea}{\end{eqnarray}}
\newcommand{\la}{\langle}
\newcommand{\ra}{\rangle}
\newcommand{\Mn}{ M_{\mbox{\tiny N}}}

\newcommand{\di}{ {\rm d} }
\newcommand{\btau}{{{\mbox{\boldmath$\tau$}}}}
\newcommand{\bgam}{{{\mbox{\boldmath$\gamma$}}}}
\newcommand{\bDelta}{{{\mbox{\boldmath$\Delta$}}}}
\newcommand{\bnabla}{{{\mbox{\boldmath$\nabla$}}}}

\newcommand{\fslash}[1] {{\not\! #1\,}}
\newcommand{\binomial}[2]{\left(\mbox{$\!\!{\renewcommand{\arraystretch}{0.5}
  	\begin{array}{c} #1\\ #2\end{array}}\!\!$}\right)}
\newcommand{\singlesum}[1]{\sum_{\renewcommand{\arraystretch}{0.3}
  \begin{array}{l}\scriptscriptstyle{#1}\end{array}}}

\newcommand{\doublesumUp}[3]{\sum_{\renewcommand{\arraystretch}{0.6}
  \begin{array}{c}\scriptscriptstyle #1\\ \scriptscriptstyle #2\end{array}}
  ^{\scriptscriptstyle{#3}}}
\newcommand{\doublelim}[2]{{\lim_{\renewcommand{\arraystretch}{0.5}
  \begin{array}{l}\scriptscriptstyle #1\\ \scriptscriptstyle #2\end{array}}
  \!\!}}
\newcommand{\triplelim}[3]{{\lim_{\renewcommand{\arraystretch}{0.5}
  \begin{array}{c}\scriptscriptstyle {#1} \\ \scriptscriptstyle {#2} \\
  \scriptscriptstyle {#3}\end{array}} \!\!}}

\begin{document}
\title{	Polynomiality of helicity off-forward distribution functions \\
 	in the chiral quark-soliton model}
\author{P.~Schweitzer$^a$, M.~Colli$^a$, S.~Boffi$^{a,b}$\\
\footnotesize\it $^a$ 	Dipartimento di Fisica Nucleare e Teorica, 
			Universit\`a degli Studi di Pavia, I-27100 Pavia, 
			Italy\\
\footnotesize\it $^b$ 	Istituto Nazionale di Fisica Nucleare, 
			Sezione di Pavia, I-27100 Pavia, Italy}
\date{March, 2003}
\maketitle
\begin{abstract}
	The polynomiality condition -- i.e.\  the property that
	Mellin moments of off-forward distribution functions are 
	even polynomials in the skewedness parameter $\xi$ --
	is a demanding check of consistency for model approaches.
	We demonstrate that the helicity off-forward distribution functions 
	in the chiral quark-soliton model satisfy the polynomiality property.
	The proof contributes to the demonstration that the description of 
	off-forward distribution functions in the model is consistent.\\
	$\phantom{X}$
\end{abstract}
%
\section{Introduction}\label{sect-1-introduction}

Off-forward distribution functions (OFDFs) of partons \cite{first-OFDF} 
(see \cite{Ji:1998pc,Radyushkin:2000uy,Goeke:2001tz} for recent reviews)
are a rich source of new information on the internal nucleon structure
\cite{interpret}.
OFDFs can be accessed in a variety of hard exclusive processes 
\cite{Vanderhaeghen:2002ax} on which recently first data became
available \cite{experiments}.
Many efforts have been devoted to predict and/or understand these
experimental results \cite{efforts}, which -- at these early stages --
necessarily are based on physical intuition, educated guesses and model
studies.
Important insights into non-perturbative aspects of OFDFs are due
to studies in the bag model \cite{Ji:1997gm} and the chiral quark-soliton 
model ($\chi$QSM) \cite{Petrov:1998kf,Penttinen:1999th,Schweitzer:2002nm}.
More recently also calculations in a light-front Hamiltonian approach 
\cite{Mukherjee:2002pq} and in a constituent model \cite{Boffi:2002yy}
have been reported.

The $\chi$QSM is based on an effective relativistic quantum field-theory
which was derived from the instanton model of the QCD-vacuum. 
The model describes in the limit of a large number of colours $N_c$
a large variety of nucleonic properties -- among others form factors 
\cite{Christov:1995vm} and (anti)quark distribution functions 
\cite{Diakonov:1996sr} -- typically within $(10-30)\%$ without 
adjustable parameters.
One of the virtues of the $\chi$QSM is its field-theoretical character, 
which ensures the theoretical consistency of the approach.
E.g., in Ref.~\cite{Diakonov:1996sr} it was proven that the quark 
and antiquark distribution functions computed in the model satisfy 
all general QCD requirements (sum rules, inequalities, etc.).
With the same rigour it was shown that the $\chi$QSM expressions for OFDFs 
reduce to usual parton distributions in the forward limit, and that their 
first moments yield form factors \cite{Petrov:1998kf,Penttinen:1999th}.

An important property of OFDFs which puts a demanding check of the 
consistency of a model approach is the polynomiality property \cite{Ji:1998pc}
(another equally important and demanding property is positivity, 
see \cite{positivity} and references therein).
In QCD it follows from hermiticity, time-reversal, parity and 
Lorentz-invariance that the $m^{\rm th}$ moment in $x$ of an OFDF 
is at $t=0$ an even polynomial in the skewedness variable $\xi$ 
(we use the notation of \cite{Ji:1998pc,Goeke:2001tz}).
The degree of the polynomial is less than or equal to $m$ in the case of the 
unpolarized OFDFs $H^q(x,\xi,t)$ and $E^q(x,\xi,t)$, and $(m-1)$ in the case 
of the helicity OFDFs $\tilde H^q(x,\xi,t)$ and $\tilde E^q(x,\xi,t)$.

The effective low energy theory underlying the $\chi$QSM is hermitian, 
time-reversal, parity and Lorentz-invariant -- like QCD -- and so one 
can expect on general grounds that the OFDFs in the $\chi$QSM satisfy 
the polynomiality property.
In the model expressions, however, the manifestation of these properties 
becomes obscure and it is necessary to check explicitly the polynomiality
property. In \cite{Petrov:1998kf,Penttinen:1999th} it was checked 
numerically that the unpolarized and helicity OFDFs in the model 
satisfy the polynomiality property.
Of course, numerical checks have limitations due to numerical accuracy,
and in \cite{Petrov:1998kf,Penttinen:1999th} the polynomiality property
was practically verified for the lowest moments.
However, in \cite{Schweitzer:2002nm} it was shown explicitly that 
the unpolarized OFDFs in the $\chi$QSM studied in \cite{Petrov:1998kf}
satisfy the polynomiality property for arbitrary moments.

In this note we generalize the methods of Ref.~\cite{Schweitzer:2002nm} to the
case of helicity OFDFs, and we explicitly prove that the model expressions 
derived in \cite{Penttinen:1999th} satisfy the polynomiality property.
The proof presented here contributes to the demonstration, that the 
description of OFDFs in the $\chi$QSM is theoretically consistent. 
The note is organized as follows. 
In Sec.~\ref{sect-2-model} a brief introduction to the $\chi$QSM is given. 
In Sec.~\ref{sect-3-OFDF-in-model} the model expressions for the helicity
OFDFs are presented. The proof of polynomiality is given 
in Secs.~\ref{sect-4-proof-H} for the non-spin-flip and 
     in  \ref{sect-5-proof-E} for the spin-flip helicity OFDF. Finally, 
in Sec.~\ref{sect-7-conclusions} we summarize our results and conclude.
The Appendix contains a discussion of the forward limit of 
$(\tilde{H}^u-\tilde{H}^d)(x,\xi,t)$.

\section{The chiral quark-soliton model \boldmath ($\chi$QSM)}
\label{sect-2-model}

The $\chi$QSM \cite{Diakonov:1987ty} is based on the principles of chiral 
symmetry breaking and the limit of a large number of colours $N_c$.
The effective chiral relativistic quantum field theory underlying 
the $\chi$QSM is given by the partition function 
\cite{Diakonov:tw,Dhar:gh,Diakonov:1983hh}
\be
	Z_{\rm eff} = \int{\cal D}\psi\,{\cal D}\bar{\psi}\,{\cal D}U\;
	\exp\Biggl(i\int\di^4x\;\bar{\psi}\,
	(i\fslash{\partial}-M\,U^{\gamma_5})\psi\Biggr) \;\;,\;\;\;\; 
	U^{\gamma_5} = e^{i\gamma_5\tau^a\pi^a} , \label{eff-theory} \ee
where $M$ denotes the dynamical quark mass which is due to spontaneous 
breakdown of chiral symmetry and $U=\exp(i\tau^a\pi^a)$ is the $SU(2)$ 
chiral pion field.
The effective theory (\ref{eff-theory}) was derived from the instanton model 
of the QCD vacuum \cite{Diakonov:1983hh}, which provides a mechanism of 
dynamical chiral symmetry breaking, and is valid at low energies 
below a scale of about $\rho_{\rm av}^{-1}\approx 600\,{\rm MeV}$ where 
$\rho_{\rm av}$ is the average instanton size.

In the large-$N_c$ limit the nucleon can be viewed as a classical soliton 
of the pion field \cite{Witten:tx}. 
The $\chi$QSM provides a realization of this idea.
In practice the large-$N_c$ limit allows to solve the path integral over pion 
field configurations in (\ref{eff-theory}) in the saddle-point approximation 
\cite{Diakonov:1987ty}.
In leading order of the large-$N_c$ expansion the pion field is static,
and one can determine the spectrum of the effective one-particle Hamiltonian 
of the theory (\ref{eff-theory})
\be
	\hat{H}_{\rm eff}|n\ra = E_n |n\ra \;\;,\;\;\;\; 
	\hat{H}_{\rm eff} = -i\gamma^0\gamma^k\partial_k+\gamma^0MU^{\gamma_5}
	\;.\label{eff-Hamiltonian}\ee
The spectrum consists of an upper and a lower Dirac continuum, which are
distorted by the pion field as compared to continua of the free 
Dirac-Hamiltonian $\hat{H}_0 = -i\gamma^0\gamma^k\partial_k+\gamma^0 M$, 
and of a discrete bound state level of energy $E_{\rm lev}$ for a 
strong enough pion field of unity winding number.
By occupying the discrete level and the states of lower continuum each 
by $N_c$ quarks in an anti-symmetric colour state, one obtains a state 
with unity baryon number. 
The minimization of the soliton energy $E_{\rm sol}$ with respect to
variations of the chiral field $U$ yields the so called self-consistent
pion field $U_c$ and the mass of the nucleon
\be
	\Mn = E_{\rm sol}[U_c] = \min\limits_U E_{\rm sol}[U]\,, \;\;\;
	E_{\rm sol}[U] = \min\limits_U 
	N_c \biggl(E_{\rm lev}+\singlesum{E_n<0}(E_n-E_{n_0})\biggr) 
	\;.\label{soliton-energy}\ee 
The momentum of the nucleon and the spin and isospin quantum numbers 
are described by quantizing the zero modes of the soliton solution.
Corrections in $1/N_c$ can be included by considering time dependent pion 
field configurations \cite{Diakonov:1987ty}. The results of the $\chi$QSM 
respect all general counting rules of the large-$N_c$ phenomenology.

For the following it is important to note that the effective Hamiltonian 
$\hat{H}_{\rm eff}$ commutes with the parity operator $\hat{\Pi}$ and the 
grand-spin operator $\hat{\bf K}$, defined as
\be\label{Def:grand-spin}
	\hat{\bf K} = \hat{\bf L}+\hat{\bf S}+\hat{\bf T}\;,
\ee
where 
$\hat{\bf L}=\hat{\bf r}\times\hat{\bf p}$ and 
$\hat{\bf S} = \frac12\gamma_5\gamma^0\bgam$ and
$\hat{\bf T}=\frac12\btau$ are respectively 
the quark orbital angular momentum, spin and isospin operators.
With the maximal set of commuting operators $\hat{H}_{\rm eff}$, 
$\hat{\Pi}$, $\hat{\bf K}^2$ and ${\hat K}^3$ the single quark states 
$|n\ra$ in (\ref{eff-Hamiltonian}) are characterized by the
quantum numbers parity $\pi$ and $K$, $M$ 
\be
	|n\ra = |E_n, \, \pi, \,K, \,M \ra 
	\;.\label{states} \ee
The $\chi$QSM allows to evaluate in a parameter-free way nucleon matrix 
elements of QCD quark bilinear operators as 
\ba
&&	\mbox{\hspace{-0.5cm}}
	\la N'({\bf P'})|\bar{\psi}(z_1)[z_1,z_2]\Gamma\psi(z_2)
	|N({\bf P})\ra \nonumber\\
&&	= A^{\mbox{\tiny$\Gamma$}}_{\mbox{\tiny$NN'$}}\;
	\Mn N_c \int\!\!\di^3{\bf X}\:e^{i({\bf P'-P}){\bf X}}\, 
	\singlesum{n, \rm occ} \bar{\Phi}_n({\bf z}_1-{\bf X})\Gamma
	\Phi_n({\bf z}_2-{\bf X})\,e^{iE_n(z^0_1-z^0_2)} + \dots 
	\label{matrix-elements} \ea
where $[z_1,z_2]$ is the gauge link and the dots denote terms 
subleading in the $1/N_c$ expansion, which will not be needed here. 
In (\ref{matrix-elements}) $\Gamma$ is some Dirac and flavour matrix, 
$A^{\mbox{\tiny$\Gamma$}}_{\mbox{\tiny$NN'$}}$ a constant depending on 
$\Gamma$ and the spin and flavour quantum numbers of the nucleon state 
$|N\ra=|S_3,T_3\ra$ and $\Phi_n({\bf x}) = \la{\bf x}|n\ra$. 
The sum in Eq.~(\ref{matrix-elements}) goes over occupied levels $n$  
(i.e. $n$ with $E_n\le E_{\rm lev}$), and vacuum subtraction is 
implied for $E_n < E_{\rm lev}$ as in Eq.~(\ref{soliton-energy}).
If in QCD the left hand side of (\ref{matrix-elements}) is scale dependent
then the result in the model on right hand side corresponds to scale of 
about $600\,{\rm MeV}$.

In the way sketched in Eq.~(\ref{matrix-elements}) static nucleonic 
observables \cite{Christov:1995vm}, twist-2 quark and anti-quark 
distribution functions of the nucleon at a low normalization point 
\cite{Diakonov:1996sr} have been computed in the $\chi$QSM and found to 
agree within $(10-30)\%$ with experimental data or phenomenological
parameterizations. 
In \cite{Petrov:1998kf,Penttinen:1999th} the approach has been 
generalized to describe OFDFs at a low normalization point.

\section{Off-forward distribution functions in the \boldmath $\chi$QSM}
\label{sect-3-OFDF-in-model}

The helicity quark off-forward distribution functions are defined as
\ba
&&	\mbox{\hspace{-1cm}}
	\int\!\frac{\di\lambda}{2\pi}\,e^{i\lambda x}\la {\bf P'},S_3'|
	\bar{\psi}_q(-\lambda n/2)\,\fslash{n}\,\gamma_5
	[-\lambda n/2,\lambda n/2]
	\psi_q(\lambda n/2)|{\bf P},S_3\ra
	\nonumber\\
&&	= \tilde{H}^q(x,\xi,t)\;
	  \bar{U}({\bf P'},S_3')\fslash{n}\gamma_5 U({\bf P},S_3)
 	+ \tilde{E}^q(x,\xi,t)\;
	  \bar{U}({\bf P'},S_3')\,\frac{n_\nu\Delta^\nu\!}{2\Mn}\,\gamma_5
	  U({\bf P},S_3) 
	+ \dots  \label{def-1} \ea
where the dots denote higher-twist contributions. 
The normalization scale dependence of $\tilde{H}^q(x,\xi,t)$ 
and $\tilde{E}^q(x,\xi,t)$ is not indicated for brevity.
The light-like vector $n^\mu$, the four-momentum transfer $\Delta^\mu$, the 
skewedness parameter $\xi$ and the Mandelstam variable $t$ are defined as
\be
  	n^2 = 0  		\; ,\;\;\;	
  	n(P'+P) = 2		\; ,\;\;\;
  	\Delta^\mu = (P'-P)^\mu	\; ,\;\;\;
  	n\Delta = -2\xi		\; ,\;\;\;
	t = \Delta^2		\; .\label{def-2} \ee
In (\ref{def-1}) $x\in[-1,1]$ with the understanding that for negative 
$x$ Eq.~(\ref{def-1}) describes the respective antiquark OFDF.
In the $\chi$QSM we work in the large-$N_c$ limit where $\Mn={\cal O}(N_c)$, 
$|\Delta^i|={\cal O}(N_c^0)$ and consequently $|\Delta^0|={\cal O}(N_c^{-1})$. 
The variables $x$ and $\xi$ are of ${\cal O}(N_c^{-1})$.
Choosing the 3-axis for the light-cone space direction, 
we have in the ``large-$N_c$ kinematics'' 
\be
	n^\mu = (1,0,0,-1)/\Mn = (1,-{\bf e}^3)/\Mn \;,\;\;\;
	t     = -\bDelta^2 		\;,\;\;\;
	\xi   = -\Delta^3/(2\Mn) 	\;.\label{def-large-Nc-kinematics}\ee
Different flavour combinations of the OFDFs exhibit different behaviour
in the large-$N_c$ limit \cite{Goeke:2001tz}
\ba
	(\tilde H^u-\tilde H^d)(x,\xi,t) = N_c^2 f(N_cx,N_c\xi,t) \;, &&
	(\tilde E^u-\tilde E^d)(x,\xi,t) = N_c^4 f(N_cx,N_c\xi,t) \;, 
	\label{H-E-largeNc-large}\\
	(\tilde H^u+\tilde H^d)(x,\xi,t) = N_c\, f(N_cx,N_c\xi,t) \;, &&
	(\tilde E^u+\tilde E^d)(x,\xi,t) = N_c^3 f(N_cx,N_c\xi,t) \;. 
	\label{H-E-largeNc-small} \ea
The functions $f(u,v,t)$ in 
Eqs.~(\ref{H-E-largeNc-large},~\ref{H-E-largeNc-small}) are stable in the 
large-$N_c$ limit for fixed values of the  ${\cal O}(N_c^0)$ variables 
$u=N_cx$, $v=N_c\xi$ and $t$, and of course different for the different OFDFs.

The model expressions for the leading OFDFs in (\ref{H-E-largeNc-large})
were derived in Ref.~\cite{Penttinen:1999th} and read
\ba
	(\tilde H^u-\tilde H^d)(x,\xi,t)&=& 
	-\,(2T^3)\,\frac{2\Mn N_c}{3\,(\bDelta^{\!\perp})^2} 
	\int\!\!\di^3{\bf X}\,e^{i\bDelta{\bf X}} \singlesum{n,\rm occ} 
	\int\!\frac{\di z^0}{2\pi}\, e^{iz^0(x\Mn - E_n)}\nonumber\\
	&&\times\phantom{\biggl|}
	\Phi^{\!\ast}_n({\bf X}+{\textstyle\frac{z^0}{2}}{\bf e^3})\,
	(1+\gamma^0\gamma^3)\,\gamma_5
	\biggl(-t\,\frac{\tau^3}{2}+\xi\Mn\;\bDelta\btau\biggr)
	\Phi_n({\bf X}-{\textstyle\frac{z^0}{2}}{\bf e^3})
	\;,\label{def-Hu-Hd-model}\\
	&&\phantom{a}\nonumber\\
	(\tilde E^u-\tilde E^d)(x,\xi,t)&=&
	-\,(2T^3)\,\frac{2\Mn^2 N_c}{3\,\xi\,(\bDelta^{\!\perp})^2}
	\int\!\!\di^3{\bf X}\,e^{i\bDelta{\bf X}}
	\singlesum{n,\rm occ} 
	\int\!\frac{\di z^0}{2\pi}\, e^{iz^0(x\Mn - E_n)} \nonumber\\
	&&\times\phantom{\biggl|}
	\Phi^{\!\ast}_n({\bf X}+{\textstyle\frac{z^0}{2}}{\bf e^3})\,
	(1+\gamma^0\gamma^3)\gamma_5(\btau^\perp\bDelta^\perp)
	\Phi_n({\bf X}-{\textstyle\frac{z^0}{2}}{\bf e^3})
	\;.\label{def-Eu-Ed-model} \ea
The 3-axis singled out because of the choice in (\ref{def-large-Nc-kinematics})
and $\bDelta^{\!\perp}=(\Delta^1,\Delta^2,0)$, etc.

In Ref.~\cite{Penttinen:1999th} it was demonstrated that in the forward 
limit $(\tilde H^u-\tilde H^d)(x,\xi,t)$ in (\ref{def-Hu-Hd-model}) reduces 
to the model expression for the helicity isovector distribution function
(see also the App.~\ref{App:forward})
\be\label{forw-lim-Hu-Hd}
	\lim\limits_{\Delta^\mu\to 0}
	(\tilde H^u-\tilde H^d)(x,\xi,t) = (g_1^u-g_1^d)(x) 
\ee
and that the model expressions (\ref{def-Hu-Hd-model},~\ref{def-Eu-Ed-model}) 
are correctly normalized to the axial vector and pseudoscalar form factor,
respectively
\be\label{norm-form-factors}
	\int\limits_{-1}^1\!\!\di x \;(\tilde H^u-\tilde H^d)(x,\xi,t) = 
	(G_1^u-G_1^d)(t)
	\;\;,\;\;\;\;
	\int\limits_{-1}^1\!\!\di x \;(\tilde E^u-\tilde E^d)(x,\xi,t) = 
	(G_2^u-G_2^d)(t)
	\;\;.\ee
The purpose of this note is to demonstrate that the model expressions in  
Eqs.~(\ref{def-Hu-Hd-model},~\ref{def-Eu-Ed-model}) satisfy the polynomiality 
condition, i.e.\  
\ba
	M^{q\,(m)}_{\tilde H}(\xi,0) \equiv
	\int\limits_{-1}^1\!\!\di x\:x^{m-1}\,\tilde H^q(x,\xi,0)
	= h^{q\,(m)}_0 + h^{q\,(m)}_2 \xi^2 + \dots + 
	    \cases{h^{q\,(m)}_{m-2}\xi^{m-2}\!\!\!\!\! & for $m$ even\cr
		   h^{q\,(m)}_{m-1}\xi^{m-1}\!\!\!\!\! & for $m$ odd,} &&
	\label{def-polynom-Hu-Hd}\\
	M^{q\,(m)}_{\tilde E}(\xi,0) \equiv
	\int\limits_{-1}^1\!\!\di x\:x^{m-1}\,\tilde E^q(x,\xi,0)
	= e^{q\,(m)}_0 + e^{q\,(m)}_2 \xi^2 + \dots + 
	    \cases{e^{q\,(m)}_{m-2}\xi^{m-2}\!\!\! & for $m$ even\cr
		   e^{q\,(m)}_{m-1}\xi^{m-1}\!\!\! & for $m$ odd.}&&
	\label{def-polynom-Eu-Ed} \ea
In Ref.~\cite{Penttinen:1999th} $(\tilde H^u-\tilde H^d)(x,\xi,t)$ and 
$(\tilde E^u-\tilde E^d)(x,\xi,t)$ have been computed as functions of 
(physical values of) $x$, $\xi$ and $t$, and the polynomiality conditions, 
Eqs.~(\ref{def-polynom-Hu-Hd},~\ref{def-polynom-Eu-Ed}), have been checked
by taking (numerically) moments $M^{(m)}_{\tilde H}(\xi,t)$, 
$M^{(m)}_{\tilde E}(\xi,t)$ and extrapolating (numerically) 
to the unphysical point $t=0$.
Of course, this only allows to check the polynomiality condition 
(for low moments) within the numerical accuracy. 
In the next two sections strict proofs will be given that all moments 
of the model expressions (\ref{def-Hu-Hd-model},~\ref{def-Eu-Ed-model}) 
satisfy the polynomiality property.

\section{Proof of polynomiality for \boldmath 
$(\tilde H^u-\tilde H^d)(x,\xi,t)$}
\label{sect-4-proof-H}

Let us denote by $M^{(m)}_{\tilde H}(\xi,t)$ the $m^{\rm th}$ moment 
of $(\tilde H^u-\tilde H^d)(x,\xi,t)$ in Eq.~(\ref{def-Hu-Hd-model}).
It is given by ({\it cf.} App.~A of \cite{Schweitzer:2002nm})
\ba
	M^{(m)}_{\tilde H}(\xi,t)
	&=& -\,
	\frac{(2T^3)\,2N_c}{3(\bDelta^\perp)^2 \Mn^{m-1}} 
	\singlesum{n,\rm occ}
	\sum\limits_{k=0}^{m-1}\binomial{m-1}{k}
	\frac{E_n^{m-1-k}}{2^k}\,
	\sum\limits_{j=0}^k \binomial{k}{j}\;\nonumber\\
	&&\times
	\la n|(1+\gamma^0\gamma^3)\,\gamma_5
	\biggl\{-t\,\frac{\tau^3}{2}+\xi\Mn\;\bDelta\btau\biggr\}
	(\hat{p}^3)^{j}\exp(i\bDelta\hat{\bf X})\, (\hat{p}^3)^{k-j}|n\ra 
	\;,\label{Hu-Hd-model-mom1}\ea
where $\hat{\bf p}$ and $\hat{\bf X}$ mean the free-momentum 
operator and the position operator, respectively.

The next step is to take the limit $t\to 0$ in (\ref{Hu-Hd-model-mom1}).
The moments $M^{(m)}_{\tilde H}(\xi,t)$ depend on $\xi$ and $t$ through 
$\bDelta$ in (\ref{def-large-Nc-kinematics}). 
Continuing analytically the operator $\exp(i\bDelta\hat{\bf X})$ 
to $t=0$ one obtains \cite{Schweitzer:2002nm}
\be
	\triplelim{\rm analytical}{\rm continuation}{t\to 0}
	\exp(i\bDelta\hat{\bf X}) = \sum\limits_{l_e=0}^\infty\,
	\frac{(-2i\xi\Mn|\hat{\bf X}|\,)^{l_e}}{l_e!\,}\; 
	P_{l_e}(\cos\hat{\theta}) \;,\label{Hu-Hd-an-cont} \ee
where the operator $\cos\hat{\theta}\equiv \hat{X}^3/|\hat{\bf X}|$.
The prefactor $1/(\bDelta^\perp)^2 = (-t -4\xi^2\Mn^2)^{-1}$ yields 
$(-4\xi^2\Mn^2)^{-1}$ with $t\to 0$. 
(We assume $\xi\neq0$. However, the final expressions will be
well defined also at $\xi=0$, see below.) 
Thus, the first expression in the curly brackets in (\ref{Hu-Hd-model-mom1}) 
vanishes like $t$ in the limit $t\to 0$.
Using (\ref{Hu-Hd-an-cont}) the result of analytical continuation 
$t\to 0$ of the operator $(\bDelta\btau)\exp(i\bDelta\hat{\bf X})=$
$[\btau\hat{\bf p},\exp(i\bDelta\hat{\bf X})]$ is given by
\be
	\triplelim{\rm analytical}{\rm continuation}{t\to 0}
	\btau\bDelta\exp(i\bDelta\hat{\bf X}) = 
	\sum\limits_{l_e=1}^\infty\,\frac{(-2i\xi\Mn)^{l_e}}{l_e!\,}
	[\btau\hat{\bf p},\,|\hat{\bf X}|^{l_e}P_{l_e}(\cos\hat{\theta})]
 	\;,\label{Hu-Hd-an-cont2}	
\ee
Thus we obtain for the moments (\ref{Hu-Hd-model-mom1}) at $t=0$ the result 
\ba
	M^{(m)}_{\tilde H}(\xi,0)
	&=& -\,
	\frac{(2T^3)\,N_c}{3 \Mn^{m-1}} 
	\singlesum{n,\rm occ}
	\sum\limits_{k=0}^{m-1}\binomial{m-1}{k}
	\frac{E_n^{m-1-k}}{2^k}\,
	\sum\limits_{j=0}^k \binomial{k}{j}
	\sum\limits_{l_e=1}^\infty\,\frac{(-2i\xi\Mn)^{l_e-1}}{l_e!}\;
	\nonumber\\
	&&\times
	\la n|(1+\gamma^0\gamma^3)\,\gamma_5(\hat{p}^3)^{j}
	i[\btau\hat{\bf p},\,|\hat{\bf X}|^{l_e}P_{l_e}(\cos\hat{\theta})]\,
	(\hat{p}^3)^{k-j}|n\ra \;.\label{Hu-Hd-model-mom2}
\ea
Next we use symmetries of the model to show that certain operators 
in (\ref{Hu-Hd-model-mom2}) vanish.
Consider the unitary matrix given by $G_5=\tau^2\gamma^1\gamma^3$ 
in the standard representation of $\gamma$- and $\tau$-matrices.
It has the property $G_5\gamma^\mu G_5^{-1}=(\gamma^\mu)^T$ and 
$G_5\tau^a G_5^{-1} = -(\tau^a)^T$.
In coordinate space $\hat{p}^i = -(\hat{p}^i)^T$ holds formally, 
and one finds that $G_5$ transforms in the coordinate-space representation 
the Hamiltonian $\hat{H}_{\rm eff}$ and single quark states, 
Eq.~(\ref{eff-Hamiltonian}), as 
$G_5 \hat{H}_{\rm eff}  G_5^{-1} = \hat{H}_{\rm eff}^T$ and 
$G_5 \Phi_n({\bf x}) = \Phi_n^\ast({\bf x})$ \cite{Christov:1995vm}.
Applying the $G_5$-transformation to the matrix elements in 
(\ref{Hu-Hd-model-mom2}) we find ({\it cf.} \cite{Schweitzer:2002nm}) 
\ba
	M^{(m)}_{\tilde H}(\xi,0)
	&=& -\,
	\frac{(2T^3)\,N_c}{3 \Mn^{m-1}} 
	\singlesum{n,\rm occ}
	\sum\limits_{k=0}^{m-1}\binomial{m-1}{k}
	\frac{E_n^{m-1-k}}{2^k}\,
	\sum\limits_{j=0}^k \binomial{k}{j}
	\sum\limits_{l_e=1}^\infty\,\frac{(-2i\xi\Mn)^{l_e-1}}{l_e!}\;
	\nonumber\\
	&&\times
	\la n|(\gamma^0\gamma^3)^{k+1}\,\gamma_5(\hat{p}^3)^{j}
	i[\btau\hat{\bf p},\,|\hat{\bf X}|^{l_e}P_{l_e}(\cos\hat{\theta})]\,
	(\hat{p}^3)^{k-j}|n\ra \;.\label{Hu-Hd-model-mom3}
\ea
The use of the $G_5$-transformation corresponds to exploring hermiticity 
and time reversal invariance.\footnote{
	More precisely, the $G_5$-transformation is the ``standard'' time 
	reversal operation and a simultaneous flavour-SU(2) rotation,
	and not -- as mentioned in \cite{Schweitzer:2002nm} --
	``non-standard'' time reversal. 
	For a thorough discussion of time-reversal in chiral models
	(and an explanation of the notions of standard/non-standard)
	see Ref.~\cite{Pobylitsa:2002fr} (and references therein).}
Note that $(\gamma^0\gamma^3)^k$ is equal to $\gamma^0\gamma^3$ (unity)
for odd (even) $k$ and introduced in (\ref{Hu-Hd-model-mom3}) for
notational convenience.

Next we use the parity transformation $\hat{\Pi}=\gamma^0\hat{\cal P}$,
where $\hat{\cal P} \hat{\bf X} \hat{\cal P}^{-1} =  -\hat{\bf X}$
and $\hat{\cal P} \hat{\bf p} \hat{\cal P}^{-1} =  -\hat{\bf p}$, which acts 
on the single quark states (\ref{states}) as $\hat{\Pi} |n\ra = \pi|n\ra$. 
Applying the parity transformation in (\ref{Hu-Hd-model-mom3}) we obtain
\ba
	M^{(m)}_{\tilde H}(\xi,0)
	&=& -\,
	\frac{(2T^3)\,N_c}{3 \Mn^{m-1}} 
	\singlesum{n,\rm occ}
	\sum\limits_{k=0}^{m-1}\binomial{m-1}{k}
	\frac{E_n^{m-1-k}}{2^k}\,
	\sum\limits_{j=0}^k \binomial{k}{j}
	\doublesumUp{l_e=1}{l\,\rm odd}{\infty}
	\frac{(-2i\xi\Mn)^{l_e-1}}{l_e!}\;
	\nonumber\\
	&&\times
	\la n|(\gamma^0\gamma^3)^{k+1}\,\gamma_5(\hat{p}^3)^{j}
	i[\btau\hat{\bf p},\,|\hat{\bf X}|^{l_e}P_{l_e}(\cos\hat{\theta})]\,
	(\hat{p}^3)^{k-j}|n\ra \;.\label{Hu-Hd-model-mom4} \ea
Thus the moments of $M^{(m)}_{\tilde H}(\xi,0)$ contain only even powers
of $\xi$ and what remains to be done is to demonstrate that the
infinite series over $l_e$ in (\ref{Hu-Hd-model-mom4}) terminates 
at an appropriate $l_e^{\rm max}$ such that the polynomiality condition 
(\ref{def-polynom-Hu-Hd}) holds.

For that we observe that the operators in the matrix elements in 
(\ref{Hu-Hd-model-mom4}) transform as irreducible tensor operators 
$\hat{T}^L_{\!M}$ of rank $L$ and $M=0$ under simultaneous rotations 
in space and isospin-space. More precisely, with the unitary operator 
$\hat{U}({\bf n},\alpha) = \exp(-i\,\alpha\,{\bf n}\cdot\hat{\bf K})$ 
-- where the axis ${\bf n}$ and angle $\alpha$ characterize the rotation
and $\hat{\bf K}$ as given in (\ref{Def:grand-spin})  -- irreducible tensor 
operators are defined as
\be\label{Def:irr-tens-T1}
	\hat{U}({\bf n},\alpha) \,\hat{T}^{L}_M\, \hat{U}^\dag({\bf n},\alpha) 
	= \sum\limits_{M'=-L}^L D^{(L)}_{M'M}({\bf n},\alpha)\;\hat{T}^{L}_{M'}
 	\; . \ee
Under (\ref{Def:irr-tens-T1}) $\gamma_5$, $|\hat{\bf X}|$, $\btau\hat{\bf p}$ 
transform as rank zero, $\gamma^0\gamma^3$, $\hat{p}^3$ as rank 1, 
and $P_{l_e}(\cos\hat{\theta})$ as rank $l_e$ operators 
({\it cf}.\ \cite{Schweitzer:2002nm}).
The product of several irreducible tensor operators is generally not an 
irreducible tensor operator but it can be decomposed into a sum of such.
Considering the quantum numbers of the single quark states $|n\ra$ in 
Eq.~(\ref{states}) we see that (\ref{Hu-Hd-model-mom4}) is a trace\footnote{
	$\sum_{\rm occ}=\sum_{E_n\le E_{\rm lev},\,\pi,K,M}$ is a sum 
	over matrix elements diagonal in $K$ and $M$, i.e.\  a trace.}
of those irreducible tensor operators. 
Such a trace vanishes unless the operator has rank zero \cite{Fano-Racah}. 
Thus, what we are interested in is to construct out of the operators in 
(\ref{Hu-Hd-model-mom4}) those irreducible tensor operators in which $l_e$ 
takes its largest possible value $l_e^{\rm max}$.

Consider even $k$ in (\ref{Hu-Hd-model-mom4}). 
Then there are $\gamma^0\gamma^3$ and $k$-times the operator $\hat{p}^3$
which can combine maximally to an operator of rank $(k+1)$. This rank has 
to be compensated by the rank of $P_{l_e}(\cos(\theta)$ to yield finally
an operator of rank zero. Thus $l_e^{\rm max} = k+1$.
For odd $k$ in (\ref{Hu-Hd-model-mom4}) no $\gamma^0\gamma^3$-operator
appears, and we obtain  $l_e^{\rm max} = k$.
The summation index $k$ goes from zero to $(m-1)$,
i.e. $l_e^{\rm max}$ is constrained by 
\be
	l_e^{\rm max}(m) = \cases{m-1  & for $m$ even 
                              \cr m    & for $m$ odd.}
	\label{Hu-Hd-model-mom5} \ee
Inserting this result into (\ref{Hu-Hd-model-mom4}) we obtain the desired
result
\ba
	M^{(m)}_{\tilde H}(\xi,0)
	&=& -\,
	\frac{(2T^3)\,N_c}{3 \Mn^{m-1}} 
	\singlesum{n,\rm occ}
	\sum\limits_{k=0}^{m-1}\binomial{m-1}{k}
	\frac{E_n^{m-1-k}}{2^k}\,
	\sum\limits_{j=0}^k \binomial{k}{j}
	\doublesumUp{l_e=1}{l\,\rm odd}{l_e^{\rm max}(m)}
	\frac{(-2i\xi\Mn)^{l_e-1}}{l_e!}\;
	\nonumber\\
	&&\times
	\la n|(\gamma^0\gamma^3)^{k+1}\,\gamma_5(\hat{p}^3)^{j}
	i[\btau\hat{\bf p},\,|\hat{\bf X}|^{l_e}P_{l_e}(\cos\hat{\theta})]\,
	(\hat{p}^3)^{k-j}|n\ra \;.\label{Hu-Hd-model-mom6}
\ea
By renaming $l_e\to l+1$ the above result can be written as
\ba
	M^{(m)}_{\tilde H}(\xi,0)
	&=& -\,
	\frac{(2T^3)\,N_c}{3 \Mn^{m-1}} 
	\singlesum{n,\rm occ}
	\sum\limits_{k=0}^{m-1}\binomial{m-1}{k}
	\frac{E_n^{m-1-k}}{2^k}\,
	\sum\limits_{j=0}^k \binomial{k}{j}
	\doublesumUp{l=0}{l\,\rm even}{m-1}
	\frac{(-2i\xi\Mn)^{l}}{(l+1)!}\;
	\nonumber\\
	&&\times
	\la n|(\gamma^0\gamma^3)^{k+1}\,\gamma_5(\hat{p}^3)^{j}\,
	i[\btau\hat{\bf p},\,|\hat{\bf X}|^{l+1}P_{l+1}(\cos\hat{\theta})]\,
	(\hat{p}^3)^{k-j}|n\ra \;.\label{Hu-Hd-model-mom6a}
\ea

\section{Proof of polynomiality for 
	\boldmath $(\tilde E^u-\tilde E^d)(x,\xi,t)$}
\label{sect-5-proof-E}

The $m^{\rm th}$ moment $M^{(m)}_{\tilde E}(\xi,t)$ of 
$(\tilde E^u-\tilde E^d)(x,\xi,t)$ in Eq.~(\ref{def-Eu-Ed-model})
is given by ({\it cf.} \cite{Schweitzer:2002nm})
\ba
	M^{(m)}_{\tilde E}(\xi,t)
	&=& -\,
	\frac{(2T^3)\,2N_c}{3\,\xi\,(\bDelta^\perp)^2\Mn^{m-2}} 
	\singlesum{n,\rm occ}
	\sum\limits_{k=0}^{m-1}\binomial{m-1}{k}
	\frac{E_n^{m-1-k}}{2^k}\,
	\sum\limits_{j=0}^k \binomial{k}{j}\;\nonumber\\
	&&\times
	\la n|(1+\gamma^0\gamma^3)\,\gamma_5(\bDelta^\perp\btau^\perp)
	(\hat{p}^3)^{j}\exp(i\bDelta\hat{\bf X})\, (\hat{p}^3)^{k-j}|n\ra 
	\;.\label{Eu-Ed-model-mom1}\ea
In order to continue analytically to $t\to 0$ we can use the result
in (\ref{Hu-Hd-an-cont2}) (with the difference that here the sum 
starts with $l=2$ (since for $l=1$ we have 
$i[\hat{\bf p}^\perp,|\hat{\bf X}|P_1(\cos\hat{\theta})]=$
$\bnabla^\perp \hat{X}^3=0$), i.e.\  
\ba
	M^{(m)}_{\tilde E}(\xi,0)
	&=& 
	\frac{(2T^3)\,4N_c}{3\,\Mn^{m-1}} 
	\singlesum{n,\rm occ}
	\sum\limits_{k=0}^{m-1}\binomial{m-1}{k}
	\frac{E_n^{m-1-k}}{2^k}\,
	\sum\limits_{j=0}^k \binomial{k}{j}\;\sum\limits_{l_e=2}^\infty\,
	\frac{(-2i\xi\Mn)^{l_e-3}}{l_e!\,} \nonumber\\
	&&\times
	\la n|(1+\gamma^0\gamma^3)\,\gamma_5\, (\hat{p}^3)^{j}\, 
	i[\btau^\perp\hat{\bf p}^\perp,\,|\hat{\bf X}|^{l_e}
	P_{l_e}(\cos\hat{\theta})]
	\, (\hat{p}^3)^{k-j}|n\ra 
	\;.\label{Eu-Ed-model-mom2}\ea
(Again we assume carefully $\xi\neq 0$ in this intermediate step.)
The only difference -- concerning the involved operators -- between 
(\ref{Eu-Ed-model-mom2}) and (\ref{Hu-Hd-model-mom2}) is that 
$\btau^\perp\hat{\bf p}^\perp$ appears instead of $\btau\hat{\bf p}$.
This difference is not relevant if we consider the transformation properties
of the matrix elements under $G_5$ symmetry and the parity transformations,
i.e.\  we obtain with the same steps as in the previous section 
\ba
	M^{(m)}_{\tilde E}(\xi,0)
	&=& 
	\frac{(2T^3)\,4N_c}{3\,\Mn^{m-1}} 
	\singlesum{n,\rm occ}
	\sum\limits_{k=0}^{m-1}\binomial{m-1}{k}
	\frac{E_n^{m-1-k}}{2^k}\,
	\sum\limits_{j=0}^k \binomial{k}{j}
	\doublesumUp{l_e=3}{l_e\,\rm odd}{\infty}
	\frac{(-2i\xi\Mn)^{l_e-3}}{l_e!\,} \nonumber\\
	&&\times
	\la n|(\gamma^0\gamma^3)^{k+1}\,\gamma_5\, (\hat{p}^3)^{j}\, 
	i[\btau^\perp\hat{\bf p}^\perp,\,|\hat{\bf X}|^{l_e}
	P_{l_e}(\cos\hat{\theta})]
	\, (\hat{p}^3)^{k-j}|n\ra 
	\;.\label{Eu-Ed-model-mom3}\ea
In the next step, however, the above-mentioned difference matters 
because $\btau^\perp\hat{\bf p}^\perp = \btau\hat{\bf p} - \tau^3\hat{p}^3$
can be decomposed into a rank zero and rank two operator
with respect to simultaneous rotations in space and flavour-space.
Apparently the rank-two piece in $\btau^\perp\hat{\bf p}^\perp$ allows for 
a larger $l_e^{\rm max}$ and we consider in the following this piece only.
Consider even $k$ in the matrix elements in (\ref{Eu-Ed-model-mom3}).
Then $\gamma^0\gamma^3$, $\btau^\perp\hat{\bf p}^\perp$ and the $k$-times 
appearing operator $\hat{p}^3$ allow to construct irreducible tensor 
operators of maximally rank $k+3$. Thus $l_e$ can take at most the 
value $k+3$. For odd $k$ there is no $\gamma^0\gamma^3$-operator and the
highest possible value of $l_e$ is $k+2$. 
Since $0\le k \le m-1$ in (\ref{Eu-Ed-model-mom3}) we obtain 
\be\label{Eu-Ed-model-mom4}
	l_e^{\rm max}(m) = \cases{m+1  & for $m$ even 
                              \cr m+2  & for $m$ odd.}
\ee
Inserting this result into (\ref{Eu-Ed-model-mom3}) and 
renaming the summation label $l_e\to l+3$ we obtain the desired result
\ba
	M^{(m)}_{\tilde E}(\xi,0)
	&=& 
	\frac{(2T^3)\,4N_c}{3\,\Mn^{m-1}} 
	\singlesum{n,\rm occ}
	\sum\limits_{k=0}^{m-1}\binomial{m-1}{k}
	\frac{E_n^{m-1-k}}{2^k}\,
	\sum\limits_{j=0}^k \binomial{k}{j}
	\doublesumUp{l=0}{l\,\rm even}{m-1}
	\frac{(-2i\xi\Mn)^{l}}{l!} \nonumber\\
	&&\times
	\la n|(\gamma^0\gamma^3)^{k+1}\,\gamma_5\, (\hat{p}^3)^{j}\, 
	i[\btau^\perp\hat{\bf p}^\perp,\,|\hat{\bf X}|^{l+3}
	P_{l+3}(\cos\hat{\theta})]
	\, (\hat{p}^3)^{k-j}|n\ra 
	\;.\label{Eu-Ed-model-mom5}\ea

\section{Summary}
\label{sect-7-conclusions}

The recently reported first measurements of deeply virtual Compton scattering
began an exciting era which will reveal novel properties of the nucleon.
At the early stage of art non-perturbative model calculations play an 
important role as a source of inspiration for phenomenological modelling 
of OFDFs.
In this context it is important to ensure the theoretical consistency which 
provides a base for the reliability of the non-perturbative model results.

In this note we have presented a study of the helicity OFDFs in the
$\chi$QSM -- in which such prominent observations have been made like the 
$D$-term \cite{D-term} in the unpolarized OFDFs \cite{Petrov:1998kf}
or the pion pole contribution in $(\tilde E^u-\tilde E^d)(x,\xi,t)$
\cite{Penttinen:1999th}.
Here we have shown explicitly that the $\chi$QSM expressions for the helicity 
OFDFs $(\tilde H^u-\tilde H^d)(x,\xi,t)$ and $(\tilde E^u-\tilde E^d)(x,\xi,t)$
satisfy the polynomiality condition, which for model approaches is one of 
the most demanding properties of OFDFs to fulfil.
The work reported here supplements the proof given in \cite{Schweitzer:2002nm}
that also the unpolarized OFDFs in the $\chi$QSM satisfy the polynomiality
condition.

This note makes a further contribution to the demonstration that the 
description of OFDFs in the $\chi$QSM is theoretically consistent, and 
helps to increase the confidence into predictions and estimates based on 
or inspired by the results from that model.

\vspace{0.5cm}
{We would like to thank M.~V.~Polyakov for discussions.
This work has partly been performed under the contract  
HPRN-CT-2000-00130 of the European Commission.}

\appendix
\section{\boldmath The forward limit of $(\tilde H^u-\tilde H^d)(x,\xi,t)$}
\label{App:forward}

In Ref.~\cite{Penttinen:1999th} it was checked that in the forward 
limit\footnote{
	More precisely in the model one has to take 
	the limit $\Delta^i\to 0$ ($i=1,\,2,\,3$) because 
	$|\Delta^0|={\cal O}(N_c^{-1}) \ll |\Delta^i|={\cal O}(N_c^0)$,
	see Eq.~(\ref{def-large-Nc-kinematics}) and the text above it.}
$\Delta^\mu\to 0$ the model expression for the helicity OFDF 
$(\tilde H^u-\tilde H^d)(x,\xi,t)$ reduces to the helicity distribution 
function $(g_1^u-g_1^d)(x)$, see Eq.~(\ref{forw-lim-Hu-Hd}). In the $\chi$QSM 
$(g_1^u-g_1^d)(x)$ is given in leading order of the large-$N_c$ limit by 
\be\label{App-0}
	(g_1^u-g_1^d)(x) = -\,(2T^3)\,\frac{\Mn N_c}{3} \singlesum{n,\rm occ} 
	\la n|(1+\gamma^0\gamma^3)\,\gamma_5\tau^3\delta(x\Mn- E_n-\hat{p}^3)
	|n\ra\;.\ee
Here we will repeat this check and explicitly demonstrate that the correct 
forward limit is obtained -- irrespective the way one takes it.
In particular we shall see that
\ba
	(g_1^u-g_1^d)(x) 
	&=& \lim\limits_{\Delta^i\to 0}
	    (\tilde H^u-\tilde H^d)(x,\xi,t)	      	\label{Delta-i}\\
	&=& \lim\limits_{t\to 0}\biggl[\doublelim{\xi\to 0}{t\neq 0}
	    (\tilde H^u-\tilde H^d)(x,\xi,t)\biggr]    	\label{first-xi}\\
	&=& \lim\limits_{\xi\to 0}\biggl[\doublelim{t \to 0}{\xi\neq 0}
	    (\tilde H^u-\tilde H^d)(x,\xi,t)\biggr]\;. 	\label{first-t}\ea
Note that Eq.~(\ref{first-t}) provides a cross check for 
model expression analytically continued to $t=0$. 

\paragraph{\boldmath Limit $\Delta^i\to 0$.}
Let us introduce 
$\bDelta_\epsilon=\epsilon(\sin\alpha\cos\beta,\sin\alpha\sin\beta,\cos\alpha)$
where $\epsilon>0$ and $\alpha\in[0,2\pi]$ and $\beta\in[0,\pi]$ are arbitrary 
angles. (However, we exclude that $\alpha$ is an integer multiple of $\pi$. 
This restriction can be droped in the final expression.) Then 
$t_\epsilon=-\epsilon^2$, $(\bDelta^\perp_\epsilon)^2=\epsilon^2\sin^2\alpha$,
$2\xi_\epsilon\Mn = -\epsilon\,\cos\alpha$. 
The limit $\Delta^i\to 0$ will be taken by letting $\epsilon\to 0$. With the 
function $f(\alpha,\beta,{\bf X})=\bDelta_\epsilon {\bf X}/\epsilon$ which 
does not depend on $\epsilon$ we obtain from (\ref{def-Hu-Hd-model})
\ba
&&	(\tilde H^u-\tilde H^d)(x,\xi_\epsilon,t_\epsilon)= 
	-\,(2T^3)\,\frac{\Mn N_c}{3} \int\!\!\di^3{\bf X}\,
	e^{i\epsilon|{\bf X}|f(\alpha,\beta,{\bf X})} \singlesum{n,\rm occ} 
	\int\!\frac{\di z^0}{2\pi}\,e^{iz^0(x\Mn - E_n)}\nonumber\\
&&	\;\;\;\times\phantom{\biggl|}
	\Phi^{\!\ast}_n({\bf X}+{\textstyle\frac{z^0}{2}}{\bf e^3})\,
	(1+\gamma^0\gamma^3)\,\gamma_5 \biggl(\tau^3
	-(\tau^1\cos\beta+\tau^2\sin\beta)\cot\alpha\biggr)
	\Phi_n({\bf X}-{\textstyle\frac{z^0}{2}}{\bf e^3})
	\;.\label{App-1-a} \ea
Taking the limit $\epsilon\to 0$ in (\ref{App-1-a})
and using $\Phi_n({\bf X}-{\textstyle\frac{z^0}{2}}{\bf e^3}) = $ 
$\la{\bf X}|\exp(i\frac{z^0}{2}\hat{p}^3)|n\ra$ (analogue for 
$\Phi^{\!\ast}_n({\bf X}+{\textstyle\frac{z^0}{2}}{\bf e^3})$) 
and $\int\di^3{\bf X}|{\bf X}\ra\la{\bf X}| = 1$ 
and integrating over $z^0$ we obtain
\ba
	(\tilde H^u-\tilde H^d)(x,0,0) 
	&=& -\,(2T^3)\,\frac{\Mn N_c}{3} \singlesum{n,\rm occ} 
	\la n| (1+\gamma^0\gamma^3)\,\gamma_5\delta(x\Mn- E_n-\hat{p}^3)
	\nonumber\\	
	&&\times \biggl\{\tau^3
	-(\tau^1\cos\beta+\tau^2\sin\beta)\cot\alpha\biggr\}|n\ra
	\;.\label{App-1-b} \ea
Let us now consider a simultaneous rotation in space and isospin space
about the 3-axis around the angle $\pi$. The net effect of this rotation is to
leave the contribution of $\tau^3$ in the curly brackets in (\ref{App-1-b}) 
invariant but to change the signs of the contributions of $\tau^1$ and 
$\tau^2$. This shows that these contributions are strictly zero and 
that (\ref{App-1-b}) coincides with the model expression (\ref{App-0})
for $(g_1^u-g_1^d)(x)$ -- which verifies Eq.~(\ref{Delta-i}).

\paragraph{\boldmath Limit $\xi\to 0$ for $t\neq 0$ and subsequent $t\to 0$.}
Taking $\xi\to 0$ in Eq.~(\ref{def-Hu-Hd-model}) we obtain
\ba
	(\tilde H^u-\tilde H^d)(x,0,t)= 
	-\,(2T^3)\,\frac{\Mn N_c}{3} 
	\singlesum{n,\rm occ} 
	\la n|(1+\gamma^0\gamma^3)\,\gamma_5\tau^3\,
	\delta(x\Mn - E_n-\hat{p}^3)\,e^{i\bDelta^\perp\hat{\bf X}^\perp}|n\ra
	\;.\label{App-2-a}\ea
In (\ref{App-2-a}) we have used $t=-(\Delta^\perp)^2$ for $\xi=0$
and performed steps similar to those leading to Eq.~(\ref{App-1-b})  
(benefiting from $[\hat{p}^3,\,\hat{\bf X}^\perp]=0$).
The right-hand-side of (\ref{App-2-a}) depends on $t$ through $\bDelta^\perp$.
We introduce $\Delta^\perp_\epsilon=\epsilon(\sin\alpha,\,\cos\alpha)$ 
with an arbitrary angle $\alpha$ such that $t_\epsilon = -\epsilon^2$.
The limit $t\to 0$ can be taken by letting $\epsilon\to 0$. In this
limit (\ref{App-2-a}) goes into (\ref{App-0}) -- which verifies
Eq.~(\ref{first-xi}).

\paragraph{\boldmath Limit $t\to 0$ for $\xi\neq 0$ and subsequent $\xi\to 0$.}
We recall that $(\bDelta^{\!\perp})^2 = -t -(2\xi\Mn)^2$.
Taking first the limit $t\to 0$ we obtain upon use of (\ref{Hu-Hd-an-cont2})
\ba
	(\tilde H^u-\tilde H^d)(x,\xi,0)&=& 
	-\,(2T^3)\,\frac{\Mn N_c}{3}
	\int\!\!\di^3{\bf X}\, 
	\singlesum{n,\rm occ} 
	\int\!\frac{\di z^0}{2\pi}\, e^{iz^0(x\Mn - E_n)}	
	\sum\limits_{l_e=1}^\infty\,\frac{(-2i\xi\Mn)^{l_e-1}}{l_e!\,}
	\nonumber\\
	&&\times\phantom{\biggl|}
	\Phi^{\!\ast}_n({\bf X}+{\textstyle\frac{z^0}{2}}{\bf e^3})\,
	(1+\gamma^0\gamma^3)\,\gamma_5\,	
	i[\btau\hat{\bf p},\,|\hat{\bf X}|^{l_e}P_{l_e}(\cos\hat{\theta})]
	\Phi_n({\bf X}-{\textstyle\frac{z^0}{2}}{\bf e^3})
	\;.\label{App-3-a}\ea
For $\xi\to 0$ in (\ref{App-3-a}) only $l_e=1$ in the sum over $l_e$ 
contributes. 
Considering $i[\btau\hat{\bf p},\,|\hat{\bf X}| P_1(\cos\hat{\theta})]=\tau^3$
we recover from (\ref{App-3-a}) the expression for $(g_1^u-g_1^d)(x)$ in 
(\ref{App-0}) -- which verifies Eq.~(\ref{first-t}).



\begin{thebibliography}{99} 

\bibitem{first-OFDF}
  D.~M\"uller, D.~Robaschik, B.~Geyer, F.~M.~Dittes and J.~Ho\u{r}ej\u{s}i,
  Fortsch.\ Phys.\  {\bf 42}, 101 (1994).
  \\
  A.~V.~Radyushkin,
  Phys.\ Lett.\ B {\bf 385}, 333 (1996);
  Phys.\ Rev.\ D {\bf 56}, 5524 (1997).
  \\
  X.~D.~Ji,
  Phys.\ Rev.\ Lett.\  {\bf 78}, 610 (1997);
  Phys.\ Rev.\ D {\bf 55}, 7114 (1997).
  \\
  J.~C.~Collins, L.~Frankfurt and M.~Strikman,
  Phys.\ Rev.\ D {\bf 56}, 2982 (1997).

\bibitem{Ji:1998pc}
  X.~D.~Ji,
  J.\ Phys.\ G {\bf 24}, 1181 (1998).

\bibitem{Radyushkin:2000uy}
  A.~V.~Radyushkin,
  in {\sl At the frontier of particle physics}, ed. M.~Shifman 
  (World Scientific, Singapore, 2001), vol.~2, p.~1037
  [arXiv:hep-ph/0101225]. 

\bibitem{Goeke:2001tz}
  K.~Goeke, M.~V.~Polyakov and M.~Vanderhaeghen,
  Prog.\ Part.\ Nucl.\ Phys.\  {\bf 47}, 401 (2001).

\bibitem{interpret}
  J.~P.~Ralston and B.~Pire,
  Phys.\ Rev.\ D {\bf 66}, 111501 (2002).
  \\
  M.~Burkardt,
  Int.\ J.\ Mod.\ Phys.\ A {\bf 18}, 173 (2003).
  \\
  M.~V.~Polyakov,
  Phys.\ Lett.\ B {\bf 555}, 57 (2003).

\bibitem{Vanderhaeghen:2002ax}
  For a recent overview see: M.~Vanderhaeghen,
  Nucl.\ Phys.\ A {\bf 711}, 109 (2002).

\bibitem{experiments}
  A.~Airapetian {\it et al.}  [HERMES Collaboration],
  Phys.\ Rev.\ Lett.\  {\bf 87}, 182001 (2001).
  \\
  S.~Stepanyan {\it et al.}  [CLAS Collaboration],
  Phys.\ Rev.\ Lett.\  {\bf 87}, 182002 (2001).
\\
  C.~Adloff {\it et al.}  [H1 Collaboration],
  Phys.\ Lett.\ B {\bf 517}, 47 (2001).

\bibitem{efforts}
  M.~Vanderhaeghen, P.~A.~Guichon and M.~Guidal,
  Phys.\ Rev.\ D {\bf 60}, 094017 (1999).
\\
  N.~Kivel, M.~V.~Polyakov and M.~Vanderhaeghen,
  Phys.\ Rev.\ D {\bf 63}, 114014 (2001).
\\
  A.~V.~Belitsky, D.~M\"uller, A.~Kirchner and A.~Sch\"afer,
  Phys.\ Rev.\ D {\bf 64}, 116002 (2001).
\\
  V.~A.~Korotkov and W.~D.~Nowak,
  Eur.\ Phys.\ J.\ C {\bf 23}, 455 (2002).
\\
  A.~V.~Belitsky, D.~M\"uller and A.~Kirchner,
  Nucl.\ Phys.\ B {\bf 629}, 323 (2002).
\\
  A.~Freund and M.~F.~McDermott,
  Phys.\ Rev.\ D {\bf 65}, 074008 (2002);
  Eur.\ Phys.\ J.\ C {\bf 23}, 651 (2002).
  A.~Freund, M.~McDermott and M.~Strikman,
  Phys.\ Rev.\ D {\bf 67}, 036001 (2003).
\\
  A.~Kirchner and D.~M\"uller,
  arXiv:hep-ph/0302007.

\bibitem{Ji:1997gm}
  X.~D.~Ji, W.~Melnitchouk and X.~Song,
  Phys.\ Rev.\ D {\bf 56}, 5511 (1997).

\bibitem{Petrov:1998kf}
  V.~Y.~Petrov, P.~V.~Pobylitsa, M.~V.~Polyakov, I.~B\"ornig, K.~Goeke and 
  C.~Weiss,
  Phys.\ Rev.\ D {\bf 57}, 4325 (1998).

\bibitem{Penttinen:1999th}
  M.~Penttinen, M.~V.~Polyakov and K.~Goeke,
  Phys.\ Rev.\ D {\bf 62}, 014024 (2000).

\bibitem{Schweitzer:2002nm}
  P.~Schweitzer, S.~Boffi and M.~Radici,
  Phys.\ Rev.\ D {\bf 66}, 114004 (2002);
  Nucl.\ Phys.\ A {\bf 711}, 207 (2002).

\bibitem{Mukherjee:2002pq}
  A.~Mukherjee and M.~Vanderhaeghen,
  Phys.\ Lett.\ B {\bf 542}, 245 (2002);
  arXiv:hep-ph/0211386. 
\bibitem{Boffi:2002yy}
  S.~Boffi, B.~Pasquini and M.~Traini,
  Nucl.\ Phys.\ B {\bf 649}, 243 (2003).

\bibitem{Christov:1995vm}
  C.~V.~Christov {\it et al.},
  Prog.\ Part.\ Nucl.\ Phys.\  {\bf 37}, 91 (1996).

\bibitem{Diakonov:1996sr}
  D.~I.~Diakonov {\it et al.},
  Nucl.\ Phys.\ B {\bf 480}, 341 (1996);
  Phys.\ Rev.\ D {\bf 56}, 4069 (1997).
\\
  P.~V.~Pobylitsa and M.~V.~Polyakov,
  Phys.\ Lett.\ B {\bf 389}, 350 (1996).
\\
  P.~V.~Pobylitsa {\it et al.},
  Phys.\ Rev.\ D {\bf 59}, 034024 (1999).
\\
  M.~Wakamatsu and T.~Kubota,
  Phys.\ Rev.\ D {\bf 60}, 034020 (1999).
\\
  K.~Goeke {\it et al.},
  Acta Phys.\ Polon.\ B {\bf 32}, 1201 (2001).
\\
  P.~Schweitzer {\it et al.},
  Phys.\ Rev.\ D {\bf 64}, 034013 (2001).

\bibitem{positivity}
  P.~V.~Pobylitsa,
  Phys.\ Rev.\ D {\bf 65}, 077504 
  and
  114015 (2002);
  Phys.\ Rev.\ D {\bf 66}, 094002 (2002);
  Phys.\ Rev.\ D {\bf 67}, 034009 (2003);
  arXiv:hep-ph/0210238;
  arXiv:hep-ph/0211160.

\bibitem{Diakonov:1987ty}
  D.~I.~Diakonov, V.~Y.~Petrov and P.~V.~Pobylitsa,
  Nucl.\ Phys.\ B {\bf 306}, 809 (1988).
\\
  D.~I.~Diakonov and V.~Y.~Petrov,
  JETP Lett.\  {\bf 43} (1986) 75
  [Pisma Zh.\ Eksp.\ Teor.\ Fiz.\  {\bf 43} (1986) 57].

\bibitem{Diakonov:tw}
  D.~I.~Diakonov and M.~I.~Eides,
  JETP Lett.\  {\bf 38}, 433 (1983)
  [Pisma Zh.\ Eksp.\ Teor.\ Fiz.\  {\bf 38}, 358 (1983)].

\bibitem{Dhar:gh}
  A.~Dhar, R.~Shankar and S.~R.~Wadia,
  Phys.\ Rev.\ D {\bf 31}, 3256 (1985).

\bibitem{Diakonov:1983hh}
  D.~I.~Diakonov and V.~Y.~Petrov,
  Nucl.\ Phys.\ B {\bf 245}, 259 (1984);
  Nucl.\ Phys.\ B {\bf 272}, 457 (1986) 457.



\bibitem{Witten:tx}
  E.~Witten,
  Nucl.\ Phys.\ B {\bf 223}, 433 (1983).

\bibitem{Pobylitsa:2002fr}
  P.~V.~Pobylitsa,
  arXiv:hep-ph/0212027.

\bibitem{Fano-Racah}
  U.~Fano and G.~Racah, {\sl Irreducible tensorial sets}
  (Academic Press, New York, 1959), pp.~79.

\bibitem{D-term}
  M.~V.~Polyakov and C.~Weiss,
  Phys.\ Rev.\ D {\bf 60}, 114017 (1999).
\\
  O.~V.~Teryaev,
  Phys.\ Lett.\ B {\bf 510}, 125 (2001).

\end{thebibliography}
\end{document}